\documentclass[reprint,twocolumn,showpacs,superscriptaddress,longbibliography,email,floatfix,aps,prl]{revtex4-2}

\usepackage{header}
\usepackage{tikz_helper}
\usepackage{xcolor} 

\begin{document}
\title{Revealing emergent many-body phenomena by analyzing large-scale space-time records of monitored quantum systems}

\author{Marcel Cech}
\affiliation{Institut f\"ur Theoretische Physik and Center for Integrated Quantum Science and Technology, Universit\"at T\"ubingen, Auf der Morgenstelle 14, 72076 T\"ubingen, Germany}
\author{Cecilia De Fazio}
\affiliation{Institut f\"ur Theoretische Physik and Center for Integrated Quantum Science and Technology, Universit\"at T\"ubingen, Auf der Morgenstelle 14, 72076 T\"ubingen, Germany}
\author{Mar\'ia Cea}
\affiliation{Max-Plank-Institut f\"ur Quantenoptik, Hans-Kopfermann-Str. 1, D-85748 Garching, Germany}
\affiliation{Munich Center for Quantum Science and Technology (MCQST), Schellingstr. 4, D-80799 M\"unchen, Germany}
\author{Mari Carmen Ba\~nuls}
\affiliation{Max-Plank-Institut f\"ur Quantenoptik, Hans-Kopfermann-Str. 1, D-85748 Garching, Germany}
\affiliation{Munich Center for Quantum Science and Technology (MCQST), Schellingstr. 4, D-80799 M\"unchen, Germany}
\author{Igor Lesanovsky}
\affiliation{Institut f\"ur Theoretische Physik and Center for Integrated Quantum Science and Technology, Universit\"at T\"ubingen, Auf der Morgenstelle 14, 72076 T\"ubingen, Germany}
\affiliation{School of Physics and Astronomy and Centre for the Mathematics and Theoretical Physics of Quantum Non-Equilibrium Systems, The University of Nottingham, Nottingham, NG7 2RD, United Kingdom}
\author{Federico Carollo}
\affiliation{Dipartimento di Fisica, Sapienza Università di Roma, Piazzale Aldo Moro 5, 00185 Rome, Italy}
\affiliation{Centre for Fluid and Complex Systems, Coventry University, Coventry, CV1 2TT, United Kingdom}

\begin{abstract}
    Recent advances in quantum simulators permit unitary evolution interspersed with locally resolved mid-circuit measurements. This paves the way for the observation of large-scale space-time structures in quantum trajectories and opens a window for the \emph{in situ} analysis of complex dynamical processes. We demonstrate this  idea using a paradigmatic dissipative spin model, which can be implemented, e.g., on Rydberg quantum simulators. Here, already the trajectories of individual experimental runs reveal surprisingly complex statistical phenomena. In particular, we exploit free-energy functionals for trajectory ensembles to identify dynamical features reminiscent of hydrophobic behavior observed near the liquid-vapor transition in the presence of solutes in water. We show that these phenomena are observable in experiments and discuss the impact of common imperfections, such as readout errors and disordered interactions. 
\end{abstract}

\maketitle


\begin{figure}
    \centering
    \includegraphics{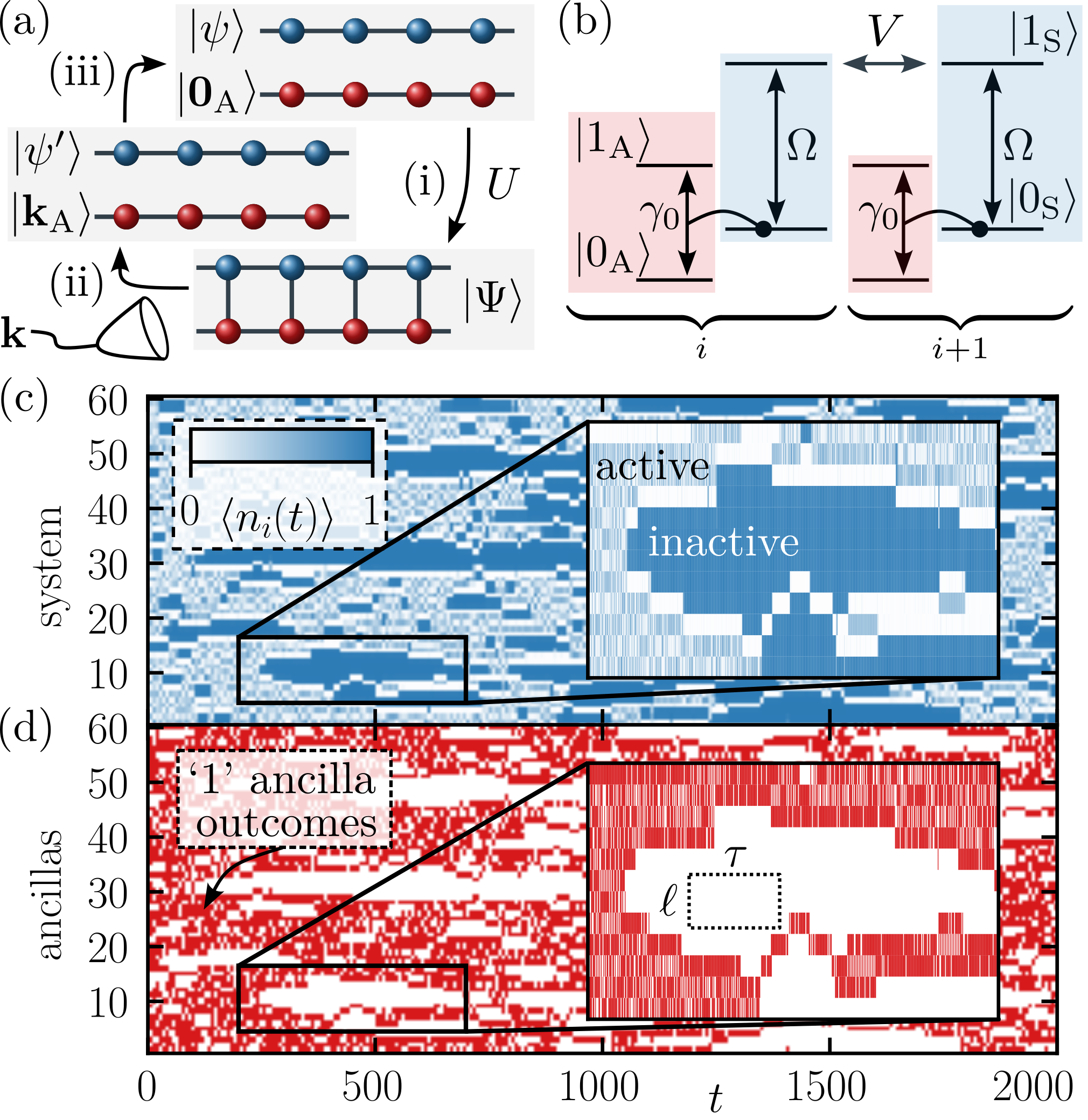}
    \caption{\textbf{Dynamical coexistence in the space-time of monitored quantum systems.} (a) The dynamics is realized as follows: (i)~the system of interest interacts with a set of ancillary degrees of freedom; (ii)~the ancillas are projectively measured,  providing the vector $\mathbf{k}$ with all binary outcomes $\mathbf{k} = (k_1, ..., k_L)$; (iii)~the ancillas are reset to their reference state $\ket{\mathbf{0}_\mathrm{A}}$. (b)~Sketch of the dynamical processes with system atoms in blue and ancillas in red [see Eq.~\eqref{eq:collision_model_hamiltonian} below].
    (c,d)~Stochastic dynamical realization of the quantum expectation of the local density of excited states  ($\ket{1_\mathrm{S}}$), $\expval{n_i(t)}$, and corresponding measurement history, for $L=60$, $\Omega \Delta t = 1.25$, $\gamma = 3\,\Omega$ and $V=5.875\,\Omega$. We observe distinct dynamical phases: an active one with rapidly varying local density and a large number of `1' measurement outcomes for the ancilla, and an inactive one with atoms  approximately frozen in the excited state and extended space-time regions characterized by `0' measurement outcomes. In one such region, we highlight a $\ell \times \tau$ cluster solely composed by `0' outcomes. Note that only panel~(d) can be obtained from a single experimental run. }
    \label{fig:fig1}
\end{figure}

\textbf{Introduction.--- }  Quantum simulators have nowadays become a versatile tool to probe both equilibrium \cite{Shao2024,Impertro2025,Schuckert2025,Haghshenas2024} and non-equilibrium \cite{Gross2017,Fauseweh2024,DarkwahOppong2022,Kim2023,Guo2024,Abanin2025,Yan2017,Yan2017a,Russ2025} many-body quantum physics.
For instance, they provide a handle to explore thermalization in closed quantum systems \cite{Kaufman2016,Zhou2022,Andersen2025} or the lack of it in the presence of ergodicity-breaking mechanisms  \cite{Schreiber2015,Choi2016,Morong2021,Guo2021}. These platforms also allow us to investigate complex many-body dynamics, e.g., stemming from Hilbert space fragmentation \cite{Scherg2021,Adler2024,Shi2024,Karch2025,Zhao2025a,Honda2025} and from the presence of quantum scars \cite{Bernien2017,Su2023,Bao2024}. 

Recent technical advances with quantum simulators permit fast high-fidelity mid-circuit readout \cite{Corcoles2021,Pino2021,Deist2022,Graham2023,Anand2024}. 
Several works, both experimental and theoretical, have exploited this capability to explore the interplay between coherent and dissipative processes as well as to engineer non-equilibrium many-body phases of matter \cite{Harrington2022,Diehl2008,Verstraete2009,Cian2019,Paviglianiti2025,Chertkov2023,Mi2024,Sun2025a,Li2025,Chen2025b}. Mid-circuit measurements further offer a way to actively intervene in the system dynamics, e.g., through feedback operations conditional on the measurement outcomes \cite{Steffen2013,Minev2019,Iqbal2024}. This turns out to be crucial, e.g., for the implementation of quantum error correction strategies \cite{Cramer2016,Krinner2022,Singh2023,Postler2024,Bluvstein2024,Acharya2024}. Moreover, as recently discovered for continuously-monitored open quantum systems \cite{Macieszczak2016a,Ilias2022,Cabot2024a,Mattes2025,Midha2025,Khan2025,Lee2025}, the statistics of mid-circuit measurements may be a resource for quantum-enhanced sensing \cite{Ilias2025} or for characterizing bath correlation functions \cite{Wang2019,Wu2024a}. 

Mid-circuit measurements are a means to manipulate quantum states or to enhance the performance of quantum devices. 
However, monitoring the measurement results of a selected subset of degrees of freedom  [cf.~Fig.~\ref{fig:fig1}(a)] also naturally yields  space-time resolved quantum trajectories. 
This information---which is directly accessible in experiments---can be analyzed to probe many-body quantum dynamics without the need of state tomography and/or postselection, merely from a standard statistical analysis [see, e.g., Fig.~\ref{fig:fig1}(c,d)]. 
In fact, the close resemblance of these space-time records to microscopic configurations of classical spin systems, see Fig.~\ref{fig:fig1}(d), motivates their analysis through  thermodynamic concepts \cite{Touchette2009,Garrahan2010,Chetrite2013,Cech2023}. While this idea has been successfully applied to study classical stochastic processes \cite{Jack2010,Chandler2010,Bodineau2012,Jack2013,Jack2015,Gutierrez2016,Garrahan2018,Buca2019,Banuls2019,Causer2020,Causer2021,Casert2021,Causer2022,Causer2022a,Sfairopoulos2025}, much less has been established in many-body quantum settings, in particular when going beyond effectively classical theories \cite{Lesanovsky2013,Valado2016,Levi2016,PerezEspigares2018,Gribben2018} and few-body cases \cite{Gribben2018,Ates2012,Lesanovsky2013a,Olmos2014,Horssen2015,Lan2018,Rose2022,ValenciaTortora2022,Yamamoto2025,Cech2025} (see however Ref.~\cite{Causer2025} for a study on a continuous-time open quantum system).  

In this paper, we explore large-scale quantum trajectories of a strongly-interacting monitored quantum system [cf.~Fig.~\ref{fig:fig1}(a,b)] and demonstrate how these can be used to unveil and characterize emergent dynamical behavior.
Using tensor-network simulations, we observe coexistence of active and inactive regions in the space-time record of ancillary measurement outcomes [see an example in Fig.~\ref{fig:fig1}(c,d)]. We introduce dynamical free-energy functions to characterize the statistics of inactive regions. This allows us to reveal phenomena akin to the {\it hydrophobic effect} and the {\it hydrophobic collapse} describing the behavior of solutes near the water-vapor phase transition. Our study highlights that modern quantum simulators with the capability to observe space-time resolved quantum trajectories open a new window for research into complex dynamical phenomena. \\


\textbf{Discrete-time quantum trajectories.--- } The setup we focus on is motivated, for instance,  by recent experiments with dual-species Rydberg atoms \cite{Anand2024}.
We consider a quantum system~(S) composed of $L$ atoms, each one coupled to an ancilla~(A), i.e., an auxiliary degree of freedom, see Fig.~\ref{fig:fig1}(a) and Refs.~\cite{Strasberg2017,Cattaneo2022,Ciccarello2022,Cech2025}. 
The latter are subject to mid-circuit measurement, which allows for indirect readout of system properties. Both atoms and ancillas consist of two-level systems with (computational) basis states $\{ \ket{0_\mathrm{S/A}}, \ket{1_\mathrm{S/A}} \}$. The dynamics effectively occurs in discrete-time and a single update is obtained as follows. The system, described by the state $\ket{\psi}$, and the ancillas, initialized in state $\ket{\mathbf{0}_\mathrm{A}} = \ket{0_\mathrm{A}}^{\otimes L}$, unitarily interact as described by the operator $U$. After the  interaction, their state reads 
\begin{align}
    \ket{\Psi} = U \left( \ket{\psi} \otimes \ket{\mathbf{0}_\mathrm{A}} \right) = \sum_{\{\mathbf{k}\}} K_\mathbf{k} \ket{\psi} \otimes \ket{\mathbf{k}_\mathrm{A}} \, , 
    \label{eq:joined_system_ancilla_state}
\end{align}
where $\mathbf{k} = (k_1, ..., k_L)$, with $k_i\in\{0,1\}$, and $K_\mathbf{k} = \bra{\mathbf{k}_\mathrm{A}} U \ket{\mathbf{0}_\mathrm{A}}$ are Kraus operators solely acting on the system \cite{Kraus1983}.  
The ancillary degrees of freedom are then projectively measured in the computational basis, which  selects a single term in the superposition appearing on the right-hand side of Eq.~\eqref{eq:joined_system_ancilla_state}. Finally, ancillas are reset to their reference state in preparation for the next update. Each  measurement outcome $\mathbf{k}$ has  probability $\pi(\mathbf{k}) = \| K_\mathbf{k} \ket{\psi} \|^2$ and is associated with the conditional system state  $\ket{\psi'} = K_\mathbf{k} \ket{\psi} / \| K_\mathbf{k} \ket{\psi} \|$. Iterating this procedure $T$ times gives rise to a stochastic realization $\ket{\psi_0} = \ket{\mathbf{0}_\mathrm{S}} \to ... \to \ket{\psi_{T}}$ and the corresponding quantum trajectory $\eta(T) = [\mathbf{k}(t)]_{t=1}^T$, encoding the space-time record of ancillary measurement outcomes. 

Concretely, we consider the unitary $U = e^{-i H_\mathrm{CM} \Delta t}$, where $\Delta t$ is the interaction time, and 
\begin{align}
    H_\mathrm{CM} = \! \Bigg( \! \Omega \sum_{i=1}^{L} \sigma_i^x + V \sum_{i=1}^{L-1} n_i n_{i+1} \! \Bigg) \!\otimes \mathds{1} +  \gamma_0 \! \sum_{i=1}^{L} P_i \otimes \sigma_i^x\, ,
    \label{eq:collision_model_hamiltonian}
\end{align}
with $\sigma^x = \ketbra{0}{1} + \mathrm{h.c}$, $n = \ketbra{1}{1}$, and $P = \ketbra{0}{0}$ \cite{Cech2025}. The left-hand side of the tensor product is for system operators while the right-hand side for ancillas [see Fig.~\ref{fig:fig1}(b)]. The first term in Eq.~\eqref{eq:collision_model_hamiltonian}  describes a chain of resonantly-driven Rydberg atoms (Rabi frequency $\Omega$) with nearest-neighbor interaction strength $V$. The second term in Eq.~\eqref{eq:collision_model_hamiltonian}  enforces a driving on the ancillas with  frequency $\gamma_0 = \sqrt{\gamma / \Delta t}$ conditional on the corresponding system atom being in $\ket{0_\mathrm{S}}$, which allows one to probe the ground-state population  \footnote{Note that, in the continuous-time limit $\Delta t \to 0$, the dynamics can be described by a Lindblad master equation with Hamiltonian $H_\mathrm{S} = \Omega \sum_{i=1}^{L} \sigma_i^x + V \sum_{i=1}^{L-1} n_i n_{i+1}$ and jump operators $J_i = \sqrt{\gamma} P_i$}. 
This model was first introduced in Ref.~\cite{Cech2025}. However, we will now explore significantly larger system sizes, which is possible by employing tensor-network techniques, see the Supplemental Material~\cite{SM}\vphantom{\cite{Fux2023,Fishman2022,Vidal2003,Vidal2004,Paeckel2019,Hatano2005,Ferris2012,Landi2023,Barthel2013,Gillman2019,FriasPerez2022a,Katira2018,Klobas2024,DeFazio2024,Lesanovsky2012a,Turner2018,Wald2025}} for details. This significant step beyond the few-body regime enables us to examine genuine many-body phenomena such as large-scale dynamical coexistence and emergent thermodynamic behavior in the space-time statistics of quantum trajectories. \\


\textbf{Quantum states vs. quantum trajectories.---} 
We are interested in characterizing dynamical features of the model described above. 
This can be achieved, for instance, by analyzing the quantum state $\ket{\psi_t}$ in single realizations of the dynamics, e.g., by considering the expectation value $\expval{n_i(t)} = \expval{n_i}{\psi_t}$ of the local Rydberg density [see an example in Fig.~\ref{fig:fig1}(c)]. For sufficiently strong interactions ($V\gg\Omega$) \cite{Cech2025}, this shows clear signatures of coexistence in terms of regions with rapid oscillations in $\expval{n_i(t)}$ and others where atoms remain stuck in the Rydberg state. 
However, for a single realization, this quantity---albeit theoretically accessible---cannot be measured in practice. 
The reason is that to estimate expectation values from experimental data one needs to collect measurements of the Rydberg density from the same stochastic realizations of the dynamics. This is not possible since the probability of observing the same trajectory twice is essentially zero. In the context of entanglement phase transitions~\footnote{The focus of this work is on the analysis of space-time records, and we do not address their application to entanglement phase transitions. For related discussions, we refer the reader to Refs.~\cite{Barratt2022, Ippoliti2024}.}\vphantom{\cite{Barratt2022, Ippoliti2024}}, this is referred to as the postselection problem \cite{Fisher2023,Passarelli2024,Garratt2024,Li2025c}. Experimentally accessible expectation values are those associated with the average state $\rho$ evolving through the Kraus map  $\mathcal{E}[\rho] = \sum_{\{ \mathbf{k} \}} K_\mathbf{k} \rho K_\mathbf{k}^\dagger$~\cite{Ciccarello2022}.
However, the information about the coexistence observed in Fig.~\ref{fig:fig1}(c) is washed out in this average state,  which eventually converges to a structureless, infinite-temperature state in the long-time limit \cite{Cech2025}.

This highlights the gain in information that is obtained from studying space-time-resolved quantum trajectories, like the one shown in Fig.~\ref{fig:fig1}(d).
Quantum trajectories are directly accessible in single experimental runs. 
No data needs to be discarded in order to gather the required empirical information and the analysis amounts to that of a standard Monte-Carlo simulation. 
In contrast, evaluating quantum expectation values of the stochastic state, or its entanglement content, would require to observe the same exact realization multiple times in experiment and thus would suffer from postselection overheads. Interestingly, however, quantum trajectories may still contain useful information about the system's dynamical behavior~\cite{Barratt2022,Ippoliti2024}. Concretely, for the model at hand, the space-time measurement records resemble the quantum expectation values $\expval{n_i(t)}$.
This becomes, e.g., apparent by the appearance of large inactive clusters---domains of `0' measurement outcomes---immersed in otherwise active regions (with a high rate of measurement outcomes `1'), that are also displayed by the Rydberg density [cf.~Fig.~\ref{fig:fig1}(c,d)].
Motivated by this and borrowing ideas from works on classical models \cite{Katira2018,Klobas2024,DeFazio2024}, we now analyze the statistics of large-scale inactive clusters in terms of thermodynamic concepts. \\


\begin{figure}
    \centering
    \includegraphics{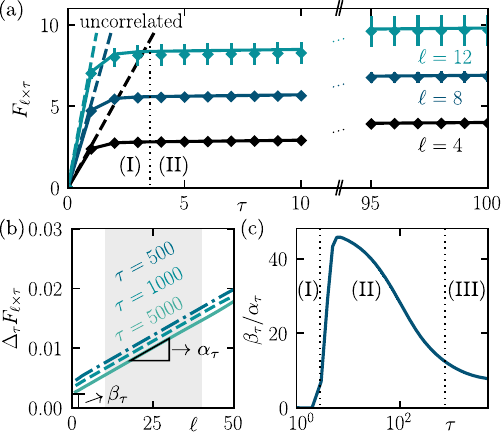}
    \caption{\textbf{Statistics of inactive space-time cluster.} (a)~Dynamical free energy (negative $\log$ of the probability) of finding an inactive cluster with fixed spatial size $\ell$ as a function of the temporal extension $\tau$. We show the empirically extracted probability from 20 quantum trajectories of $100\,000$ time steps each (symbols with errorbars representing the statistical uncertainty) and the theoretical prediction for the ensemble average  (solid lines). The dashed lines provide the dynamical free energy in the case of uncorrelated outcomes for comparison. (b,c)~Scaling behavior of the free-energy increment $\Delta_\tau F_{\ell \times \tau}$ [cf.~Eq.~\eqref{eq:scaling_function}]. For every fixed temporal extension $\tau$, we infer $\alpha_\tau$ and $\beta_\tau$ from the gray shaded region.
    }
    \label{fig:fig2}
\end{figure}

\textbf{Statistics of inactive space-time clusters.---} The coexistence of inactive and active regions within individual trajectories arises in this model for a sufficiently strong interaction $V$ (a detailed analysis on how this coexistence emerges as the Hamiltonian parameters are varied is provided in Ref.~\cite{SM}). Here, the extension of inactive dynamical regions is characterized 
through the stationary probability $p_{\ell\times\tau}$ of finding an inactive cluster of space-time size $\ell\times \tau$ in the quantum trajectory, see an example in Fig.~\ref{fig:fig1}(d). 
It is instructive to describe such a probability by means of an effective ``dynamical free energy", given by  
\begin{align}
 F_{\ell \times \tau}    = - \log p_{\ell\times\tau} \, .
    \label{eq:dynamical_free_energy} 
\end{align}
This quantity indeed allows us to understand how the probability decays in terms of an ``energetic" and ``entropic" cost of accommodating inactive clusters in quantum trajectories. As shown in Fig.~\ref{fig:fig2}(a), $F_{\ell \times \tau}$ exhibits distinct behavior as a function of $\tau$ for fixed $\ell$: (I) a regime effectively  scaling with the area of the cluster, characterized by a steep growth for $\tau \lesssim 3$, which strongly depends on $\ell$, followed by (II) a regime effectively scaling with the perimeter of the cluster, with a much slower growth for $\tau \gtrsim 3$, only weakly depending on $\ell$. The dynamical free energy thus is much smaller than that of completely uncorrelated trajectories, namely $ \ell \tau\, F_{1 \times 1}$ [see dashed lines in Fig.~\ref{fig:fig2}(a)].

Interestingly, similar crossovers from an area to a perimeter scaling have recently been observed in spin models with kinetic constraints ~\cite{Katira2018,Klobas2024,DeFazio2024}, where they have been associated with proximity to a first-order dynamical phase transition.
The underlying idea builds on the behavior of solutes near the liquid-vapor transition~\cite{Chandler2005,Willard2008,Lum1999}. Here, the free energy for encapsulating a small solute inside a vapor bubble shows a cost that scales with the size of bubble itself. For larger solutes, instead, the free energy solely scales as the interface between water and vapor.

\begin{figure}
    \centering
    \includegraphics{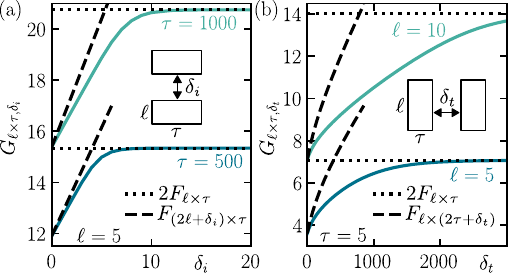}
    \caption{\textbf{Attraction between inactive clusters.}
    Dynamical free energy for two $\ell \times \tau$ inactive clusters separated by either (a) $\delta_i$ in space or (b) $\delta_t$ in time. In both panels, the dotted line represents the dynamical free energy for independent clusters, while the dashed line represents that for an inactive cluster of size $(2 \ell + \delta_i) \times \tau$ [panel~(a)] or $\ell \times (2 \tau + \delta_t)$ [panel~(b)], enclosing the two considered inactive clusters. 
    }
    \label{fig:fig3}
\end{figure}

Assuming the dynamical free energy $F_{\ell \times \tau}$ to solely consist of an area and a perimeter contribution (see the Supplemental Material~\cite{SM} for details), its increment, $\Delta_\tau F_{\ell \times \tau}$, upon extending the cluster size by one unit of time at fixed $\ell$ takes the form  
\begin{align}
    \Delta_\tau F_{\ell \times \tau}
    = F_{\ell \times (\tau+1 )} - F_{\ell \times \tau} = \alpha_\tau \ell + \beta_\tau\, .
    \label{eq:scaling_function}
\end{align}
Such a quantity is essentially related to a  temporal ``interface tension'' of the inactive cluster. 
The plot in Fig.~\ref{fig:fig2}(b) confirms the linear behavior of  Eq.~\eqref{eq:scaling_function},  which allows us to determine the area and perimeter contributions in terms of their respective coefficients $\alpha_\tau$ and $\beta_\tau$. 
For any $\alpha_\tau > 0$, the area contribution would eventually dominate in the limit of very large inactive clusters ($\ell \to \infty$).  
On the other hand, for intermediate cluster sizes, the dominant scaling behavior of the dynamical free energy can be understood by comparing the cluster's spatial extension, i.e., $\ell \approx 10$, to the fraction $\beta_\tau / \alpha_\tau$ [cf.~ Eq.~\eqref{eq:scaling_function}]. 
In Fig.~\ref{fig:fig2}(c), we highlight two area-dominated regimes, $\beta_\tau < \alpha_\tau \ell$, appearing for small and very large $\tau$, and a perimeter-dominated one, $\beta_\tau > \alpha_\tau \ell$, emerging at intermediate values of $\tau$. The first area-to-perimeter crossover is reminiscent of the hydrophobic effect close to the water-to-vapor transition. A key difference, also with respect to results for the classical models \cite{Katira2018,Klobas2024,DeFazio2024}, lies in the re-entrant area-dominated scaling for large $\tau$. The latter is essentially due to the fact that, while transitions to or from the Rydberg state are off-resonant in the presence of excited neighbors and strong interactions, they can still occur even in the middle of an otherwise inactive region. This hence allows for `1' measurement outcomes that break up the inactive cluster and provides an underlying area contribution to the dynamical free energy, which dominates at very large cluster sizes. \\


\textbf{Attraction between inactive clusters.---} Extending the discussion of inactive regions, we now investigate the dynamical free energy $G_{\ell\times \tau, \delta_i}$ ($G_{\ell\times \tau, \delta_t}$) associated with the probability of observing two identical inactive clusters of size $\ell\times\tau$ separated by a spatial (temporal) distance $\delta_{i}$ ($\delta_{t}$). This quantity is shown in Fig.~\ref{fig:fig3} and also exhibits crossover behavior. For large distances between clusters, the dynamical free energy of the two clusters saturates to a constant value, essentially given by the free energy of two independent clusters, namely $2F_{\ell \times \tau}$ (black dotted lines). For shorter distances, instead, the dynamical free energy is initially dominated by the free energy of a single larger cluster enclosing the two, i.e.,  $G_{\ell\times \tau, \delta_i} \approx F_{(2\ell + \delta_i) \times \tau}$ and $G_{\ell\times \tau, \delta_t} \approx F_{\ell \times (2 \tau + \delta_t)}$ for spatial and temporal separation, respectively (black dashed lines). As the dynamical free energy provides  a ``cost" for sustaining the configuration, its decrease for decreasing distances entails an effective attractive force between clusters, suggesting that it is likely to find clusters to merge together.  This behavior is reminiscent of the so-called ``hydrophobic collapse". 

As it can be appreciated by comparing Fig.~\ref{fig:fig3}(a) and Fig.~\ref{fig:fig3}(b), a marked difference emerges between the case of spatially or temporally separated inactive clusters. As the spatial separation increases, the dynamical free energy rapidly approaches the limit of uncorrelated inactive clusters. 
In contrast, temporal correlations between inactive clusters persist over significantly longer timescales due to the difference in temporal and spatial interface tensions (see the Supplemental Material~\cite{SM}). \\


\begin{figure}
    \centering
    \includegraphics{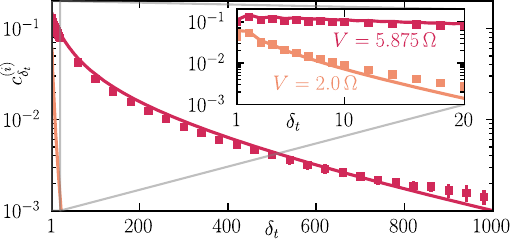}
    \caption{\textbf{Autocorrelation function of space-time records.} The solid line provides the behavior of the ensemble-average autocorrelation function, for the ideal case, as a function of the time difference $\delta_t$ [cf.~Eq.~\eqref{eq:autocorrelation}]. The symbols are results obtained from quantum trajectories considering the disordered couplings (see text) as well as a probability $p_\mathrm{err} = 2\,\%$ for measurement readout errors. The top curve is for  $V=5.875\,\Omega$ (red), where trajectories display dynamical phase coexistence. The bottom curve is for $V=2\,\Omega$ (orange), which does not display such coexistence behavior. The inset shows a magnification of the initial transient behavior. 
    }
    \label{fig:fig4}
\end{figure}

\textbf{Role of experimental imperfections.---} 
In order to explore the  robustness of the observed phenomenology to experimental imperfections, we consider, as an exemplary case, the autocorrelation function of space-time records
\begin{align}
    c_{\delta_t}^{(i)}(t) = \overline{k_i(t) k_i(t-\delta_t)} - \overline{k_i(t)} \, \overline{k_i(t - \delta_t)} \, ,
    \label{eq:autocorrelation}
\end{align}
with $\overline{f(t)}$ denoting the ensemble average of a dynamical quantity $f(t)$ over all quantum trajectories~\cite{Cech2025}.
We focus on its behavior at stationarity. Furthermore, since we expect measurements to be only marginally influenced by the position along the chain, we also average over all ancilla sites $i$ far enough from the boundaries. 
Common experimental imperfections are related to disorder in the interaction strengths between atoms and ancillas, as well as between atoms themselves. To account for this, we consider locally varying coupling constants $V$ and $\gamma$, which are uniformly distributed in a range of $\pm \Omega$. 
Additionally, we also include measurement readout errors,  characterized by a probability $p_\mathrm{err}$ of flipping the ancilla outcome. 

The autocorrelation function given by Eq.~\eqref{eq:scaling_function} is displayed in Fig.~\ref{fig:fig4}. It shows slow relaxation towards uncorrelated measurements. This slowness is in fact a result of the correlated trajectory dynamics, i.e. the coexistence of active and inactive space-time regions. The figure also compares the autocorrelation obtained through an ideal ensemble average with the one obtained by gathering statistics from quantum trajectories of more realistic dynamics. This highlights that such a quantity is robust to experimental imperfections \cite{SM}. For comparison we also show the autocorrelation function at $V = 2\,\Omega$. Here the system relaxes uniformly and correlations decay orders of magnitude faster.\\

\textbf{Outlook.---} Mid-circuit measurements in quantum simulation provide access to space-time resolved information on many-body quantum dynamics. 
Already the model analyzed in this paper shows interesting emergent dynamical behavior. Experiments with quantum simulators will allow for the study of more complex cases beyond the reach of tensor network computations. These include higher-dimensional systems and many-body dynamics, which generate more entanglement than the Rydberg density-density interaction considered here. This may lead to new collective features in trajectory thermodynamics or even to the emergence of new universality classes. \\


The code and data that support the findings of this work are available on Zenodo \cite{ZenodoData}. \\


\acknowledgments
\textbf{Acknowledgements. --- } We thank Hannes Bernien, Peter Sollich, Juan P. Garrahan and Adam Gammon-Smith for fruitful discussions. Tensor network calculations were performed using the ITensor library~\cite{Fishman2022}. We acknowledge funding from the Deutsche Forschungsgemeinschaft (DFG, German Research Foundation) under Germany's Excellence Strategy -- EXC-2111 -- 390814868, through the Research Unit FOR 5413/1, Grant No. 465199066, through the Research Unit FOR 5522/1, Grant No. 499180199, and through the state of Baden-W\"urttemberg through bwHPC and the German Research Foundation (DFG) through grant no INST 40/575-1 FUGG (JUSTUS 2 cluster). This project has also received funding from the European Union’s Horizon Europe research and innovation program under Grant Agreement No. 101046968 (BRISQ) and the Leverhulme Trust (Grant No. RPG-2024-112). C.D.F.~acknowledges support from the Alexander von Humboldt Foundation through a postdoctoral research fellowship. This work is supported by ERC grant OPEN-2QS (Grant No. 101164443, https://doi.org/10.3030/101164443).

\bibliography{biblio.bib}

\onecolumngrid
\clearpage

\setcounter{equation}{0}
\setcounter{page}{1}

\setcounter{figure}{0}
\setcounter{table}{0}
\makeatletter
\renewcommand{\theequation}{S\arabic{equation}}
\renewcommand{\thefigure}{S\arabic{figure}}
\renewcommand{\thetable}{S\arabic{table}}
\setcounter{secnumdepth}{2}

\begin{center}
{\Large SUPPLEMENTAL MATERIAL}
\end{center}
\begin{center}
\vspace{0.8cm}
{\Large Revealing emergent many-body phenomena by analyzing large-scale space-time records of monitored quantum systems}
\end{center}
\begin{center}
Marcel Cech,$^{1}$ Cecilia De Fazio,$^{1}$ Mar\'ia Cea,$^{2,3}$ Mari Carmen Ba\~nuls,$^{2,3}$ Igor Lesanovsky,$^{1,4}$ and Federico Carollo$^{5,6}$ 
\end{center}
\begin{center}
$^1${\em Institut f\"ur Theoretische Physik and Center for Integrated Quantum Science and Technology,\\ Universit\"at T\"ubingen, Auf der Morgenstelle 14, 72076 T\"ubingen, Germany}\\
$^2${\em Max-Plank-Institut f\"ur Quantenoptik, Hans-Kopfermann-Str. 1, D-85748 Garching, Germany}\\
$^3${\em Munich Center for Quantum Science and Technology (MCQST), Schellingstr. 4, D-80799 M\"unchen, Germany}
$^4${\em School of Physics and Astronomy and Centre for the Mathematics\\ and Theoretical Physics of Quantum Non-Equilibrium Systems,\\ The University of Nottingham, Nottingham, NG7 2RD, United Kingdom}\\
$^5${\em Dipartimento di Fisica, Sapienza Università di Roma, Piazzale Aldo Moro 5, 00185 Rome, Italy}\\
$^6${\em Centre for Fluid and Complex Systems, Coventry University, Coventry, CV1 2TT, United Kingdom}
\end{center}


\section{Tensor Network Methods \label{sec:TN_methods}}
In this section, we discuss the tensor network methods employed to generate the data analyzed in the main text. In Sec.~\ref{sec:QJMC}, we focus on simulations of the stochastic dynamics in terms of discrete-time quantum-jump Monte Carlo (QJMC) trajectories, which can be obtained by sampling measurement outcomes on the ancillas. In Sec.~\ref{sec:average_state_calc}, we present the appropriately conditioned average state evolution that allows us to compute quantities such as the probability of inactive space-time clusters and the autocorrelation functions. Furthermore, in Tab.~\ref{tab:TN_description}, we provide a schematic summary of the methods and parameters used for each figure in the main text.
For both instances, we explicitly include the ancillas in our tensor network ansatz (see also Ref.~\cite{Fux2023} for a similar approach). We further note that we utilize the open-source ITensor library~\cite{Fishman2022} in Julia to implement common tensor network routines like contractions and singular-value decompositions (SVDs).

\begin{table}[htbp]
    \centering
    \caption{\textbf{Tensor network specifications.} For the results presented in the main manuscript, we specify the method alongside the maximal bond dimension $\chi$ or the SVD cutoff $\epsilon$, which control the accuracy of the simulation. List of abbreviations: MPS = matrix product state, QJMC = quantum jump Monte-Carlo (see Sec.~\ref{sec:QJMC}), MPO = matrix product operator, bw-evo = backward evolution, tw-evo = evolution in two ways (see Sec.~\ref{sec:average_state_calc}).}
    \label{tab:TN_description}
    \setlength{\extrarowheight}{4pt}
    \begin{tabular}{c|c|c|c|c|c}
        Figure & Panel & Description & Method & maximal bond dimension $\chi$ & SVD cutoff $\epsilon$ \\
        \hline 
        1     & (c,d) & stochastic realization & MPS-QJMC & -     & $10^{-10}$ \\
        &       &       &       &       &  \\
        2     & (a)   & empirical $p_{\ell \times \tau}$ & MPS-QJMC & -     & $10^{-8}$ \\
                & (a-c) & $p_{\ell \times \tau}$ & MPO bw-evo & 64    & - \\
                &       &       &       &       &  \\
        3     & (a)   & $p_{\ell \times \tau}$ \& $p_{\ell \times \tau,\delta_i}$ & MPO bw-evo & 64    & - \\
                & (b)   & $p_{\ell \times \tau, \delta_t}$ & MPO tw-evo & 64    & - \\
                &       & $p_{\ell \times \tau}$ & MPO bw-evo & 64    & - \\
                &       &       &       &       &  \\
        4     & -     & empirical $c_{\delta_t}^{(i)}$ & MPS-QJMC & -     & $10^{-8}$ \\
                &       & $c_{\delta_t}^{(i)}$ & MPO tw-evo & 64    & - \\
    \end{tabular}%
\end{table}

\subsection{Discrete-time quantum-jump Monte-Carlo simulation \label{sec:QJMC}}

\begin{figure}
    \centering
    \includegraphics{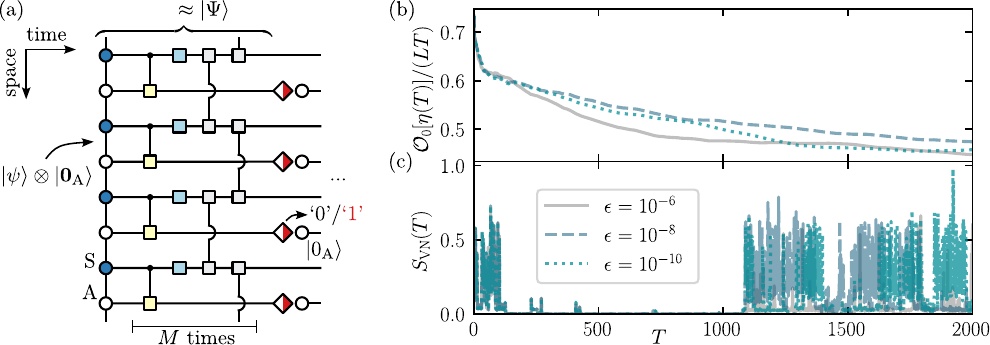}
    \caption{\textbf{Tensor network simulation of stochastic dynamics.} (a)~Tensor network diagram implementing the discrete-time quantum-jump Monte-Carlo (QJMC) evolution. The blue and white circles in the MPS diagram represent the tensors for the system and ancilla qubits, respectively. Squares denote unitary gates [see Eq.~\eqref{eq:Trotter_gates}]. Two-colored diamonds represent the projective measurements of ancillas in the computational basis $\{ \ket{0_\mathrm{A}}, \ket{1_\mathrm{A}} \}$, and are followed by the reset to $\ket{0_\mathrm{A}}$ to anticipate the next time step. 
    (b,c)~Comparison of stochastic realizations simulated with different SVD cutoffs $\epsilon$. We compare the average number of the number of `1' ancilla measurement outcomes over time [see Eq.~\eqref{eq:average_activity}] and the half-chain entanglement entropy $S_\mathrm{VN}(T)$ [see Eq.~\eqref{eq:entanglement_entropy}]. Here, $L=60$, $\Omega \Delta t = 1.25$, $V = 5.875\,\Omega$, $\gamma = 3\,\Omega$ and $M=10$.
    }
    \label{fig:fig_s1}
\end{figure}

Simulating the stochastic dynamics requires us to follow the three-step procedure depicted in Fig.~\ref{fig:fig1}(a). We here report a detailed description of how these steps are implemented with tensor networks.

At a given time $t$, the state of the system is described by the vector $\ket{\psi}$. The total system-ancillas state thus reads $\ket{\psi} \otimes \ket{\mathbf{0}_\mathrm{A}}$ and, under the subsequent unitary evolution, it becomes $\ket{\Psi} = U(\ket{\psi} \otimes \ket{\mathbf{0}_\mathrm{A}})$ [see Eq.~\eqref{eq:joined_system_ancilla_state}]. Assuming that one has an accurate Matrix Product State (MPS) representation of  $\ket{\psi} \otimes \ket{\mathbf{0}_\mathrm{A}}$, the unitary evolution, represented by step~(i) in Fig.~\ref{fig:fig1}(a), can be implemented using the time-evolving block decomposition (TEBD) algorithm \cite{Vidal2003,Vidal2004,Paeckel2019}. To this end, we approximate the unitary $U$ using a first-order Trotter decomposition \cite{Hatano2005} 
\begin{align}
    U = e^{-i H_\mathrm{CM} \Delta t} = \left(e^{-i H_\mathrm{CM} \Delta t / M}\right)^M 
    \approx 
    \left(  
        \underbrace{
        e^{-i\frac{V \Delta t}{M} \sum n_i n_{i+1}} \cdot
        e^{-i\frac{\Omega \Delta t}{M} \sum \sigma_i^x}
        }_\text{system only} \cdot
        \underbrace{
        e^{-i\frac{{\gamma_0 \Delta t}}{M} \sum P_i \otimes \sigma_i^x }
        }_\text{system-ancilla int.}
    \right)^M \, ,
    \label{eq:trotter_decomposition}
\end{align}
with $M = 10$.
The tensor network diagram in Fig.~\ref{fig:fig_s1}(a) shows this decomposition in terms of the gates:
\begin{align}
    \begingroup
    \DissGate = e^{-i\frac{{\gamma_0 \Delta t}}{M} P \otimes \sigma^x} \, , \quad \DriveGate = e^{-i\frac{\Omega \Delta t}{M} \sigma^x} \, , \quad \InteractionGate = e^{-i\frac{V \Delta t}{M} n \otimes n} \, ,
    \endgroup
    \label{eq:Trotter_gates}
\end{align}
where operators on the left-hand side of the tensor product act on the upper qubit in the graphical representation. 
Once an accurate approximation of the state $\ket{\Psi}$ has been obtained, we implement the local ancilla measurements according to Born's rule, corresponding to step~(ii) in Fig.~\ref{fig:fig1}(a). This Monte-Carlo sampling procedure, known in the context of MPS as direct (perfect) sampling algorithm \cite{Ferris2012}, returns (upon normalization) the post-measurement system-ancilla state $\ket{\psi'} \otimes \ket{\mathbf{k}_\mathrm{A}}$, where $\mathbf{k} = (k_1, ..., k_L)$ denotes the ancillas' outcome and  $\ket{\psi'} = K_\mathbf{k} \ket{\psi} / \| K_\mathbf{k} \ket{\psi} \|$ is the system's state. This procedure, combined with resetting the ancillas to the reference state $\ket{\mathbf{0}_\mathrm{A}}$, is implemented through the local, non-unitary operation $\EfficientSampling$ as shown in Fig.~\ref{fig:fig_s1}(a), and completes step (iii) in Fig.~\ref{fig:fig1}(a). 

To obtain individual realizations of the stochastic dynamics, e.g., the one shown in Fig.~\ref{fig:fig1}(c,d), we initialize the system in $\ket{\psi_0} = \ket{\mathbf{0}_\mathrm{S}}$ and iterate the above approach $T$ times.
We can control the trade-off between accuracy and computational complexity by setting a  lower limit, or cutoff, $\epsilon$ on the Schmidt coefficients that are considered, when evolving the MPS according to Fig.~\ref{fig:fig_s1}(a).
We generate the random numbers required for the direct sampling using a fixed seed. As can be seen below, this enables us to analyze the influence of the truncation more effectively, while not giving up the stochastic nature of the ancilla outcomes.\\

We now focus on two quantities: an observable on the space-time records of ancilla measurement outcomes and the entanglement entropy. For the first, we consider the average activity introduced in Ref.~\cite{Cech2025}, which quantifies the density of `1' ancilla measurement outcomes within a space-time region of size $L \times T$, namely
\begin{align}
    \frac{1}{LT} \mathcal{O}_0[\eta(T)] = \frac{1}{LT} \sum_{t=1}^T \sum_{i=1}^L k_i(t) \, .
    \label{eq:average_activity}
\end{align}
The instantaneous half-chain entanglement entropy at stroboscopic times $T$ is defined as
\begin{align}
    S_\mathrm{VN}(T) = - \Tr{\rho_T^{\mathrm{red.}} \log\left(\rho_T^{\mathrm{red.}}\right)} \, , 
    \label{eq:entanglement_entropy}
\end{align}
where we introduced the reduced density matrix:
\begin{align}
  \rho_T^{\mathrm{red.}} = \Tr_{(i > L/2)}\left\{ \ketbra{\psi_T}\right \}\,.
    \label{eq:reduced_denmatrix_hc}
\end{align}

Fig.~\ref{fig:fig_s1}(b,c) show the two quantities as functions of time.
Apart from an overall qualitative agreement, we observe that, after $T \approx 100$ time steps, both the average activity and the instantaneous half-chain entanglement entropy become sensitive to the cutoff $\epsilon$.  
We also notice that the quantities do not converge in a uniform way upon increasing the accuracy with smaller values of $\epsilon$. 
However, the instantaneous entanglement entropy does not significantly increase with smaller values of $\epsilon$ indicating that states of the stochastic realization are overall well approximated.
Discrepancies between the different curves are due to the fact that even a slight modification in the calculation of the state affects the probability of observing a certain set of ancilla measurement outcomes $\mathbf{k}$, c.f., $\pi(\mathbf{k}) = \bra{\Psi} (\mathds{1} \otimes \ketbra{\mathbf{k}_\mathrm{A}}) \ket{\Psi}$. Over a long time, this can eventually alter the outcome of a single ancilla measurement, and thereby completely change its subsequent evolution.

However, for sufficiently small values of $\epsilon$ and large enough sample sizes, we do not expect this to affect the behavior of empirical quantities such as the probability of inactive clusters and the temporal autocorrelation function. 
In practice, we find that setting the cutoff to $\epsilon = 10^{-8}$ yields accurate results for Figs.~\ref{fig:fig2}~and~\ref{fig:fig4} at relatively low computational cost.

\subsection{Ensemble averages from average state calculations \label{sec:average_state_calc}}
As a next step, we describe the average state calculations, which allow us to compute the probabilities of inactive clusters and the autocorrelation function as an ensemble average at stationarity. These quantities can be understood as observable in the space of quantum trajectories, as determined by all possible records of ancilla measurement outcomes in space and time.
This means that their long-time averages can be computed via an appropriately conditioned time evolution of the stationary state \cite{Landi2023}, here given by the fully mixed state $\rho_\mathrm{ss} = \mathds{1} / 2^L$ \cite{Cech2025}. \\

Concretely, the condition to encounter an inactive region of length $\ell$ at a given time of the dynamics can be implemented through the transfer operator $ T_\ell [\rho]$. This operator describes the average state evolution selecting only those trajectories in which at least $\ell$ central ancilla outcomes are `0', namely 
\begin{align}
    T_\ell [\rho] = \Tr_\mathrm{A} \left\{ \mathcal{P}_\ell \, U \left( \rho \otimes \ketbra{\mathbf{0}_\mathrm{A}} \right) U^\dagger \right\} \, ,
    \label{eq:transfer_operator_MPO}
\end{align}
where $\mathcal{P}_\ell =\ketbra{{0}_\mathrm{A}}^{\otimes \ell} $ is a local projector onto $\ell$ central sites of the ancillas in the zero state. Through this quantity, we can define the probability $p_{\ell\times\tau}$ of an $\ell \times \tau$-sized inactive cluster as 
\begin{align}
    p_{\ell \times \tau} =  \Tr{(T_{\ell})^{\tau} [\rho_\mathrm{ss}]} \, ,
    \label{eq:probability_inactive_regions}
\end{align}
where the required conditioned evolution is implemented by repeated application of the transfer operator $T_\ell [\rho]$ over a time interval $\tau$.
Similarly, we can also construct the conditioned evolution targeting the autocorrelation function [cf.~Eq.~\eqref{eq:autocorrelation} and Ref.~\cite{Cech2025}] and the probability of two identical inactive clusters separated in time or space, i.e., 
\begin{align}
    p_{\ell \times \tau, \delta_t} &= \Tr{(T_{\ell})^{\tau} \circ \mathcal{E}^{\delta_t} \circ (T_{\ell})^{\tau} [\rho_\mathrm{ss}]} \,, \\
    p_{\ell\times \tau, \delta_i} &= \Tr{(T_{\ell, \delta_i})^{\tau} [\rho_\mathrm{ss}]} \, ,
\end{align}
Here, the unconditioned evolution under $\mathcal{E}[\rho] = \Tr_\mathrm{A}\{ U(\rho \otimes \ketbra{0_\mathrm{A}}) U^\dagger \}$ (the accordingly conditioned transfer operator $T_{\ell, \delta_i}$) marginalizes over the $\delta_t$ ($\delta_i$) ancilla measurement outcomes between the two $\ell \times \tau$-sized inactive regions. \\

\begin{figure}[ht]
    \centering
    \includegraphics{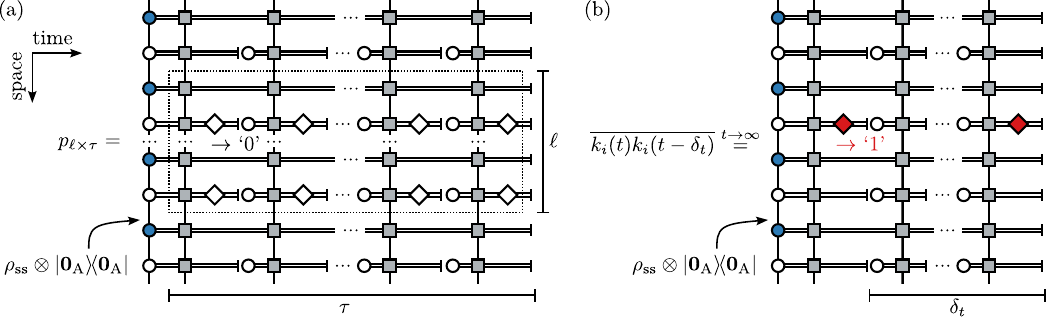}
    \caption{\textbf{Tensor network diagrams for average state calculations.} We show the tensor network diagrams corresponding to (a)~the probability $p_{\ell \times \tau}$ of an inactive space-time cluster as given by  Eqs.~(\ref{eq:transfer_operator_MPO},\ref{eq:probability_inactive_regions}) and (b)~the two-point ancilla expectation value $\overline{k_i(t) k_i(t - \delta_t)}$ entering the autocorrelation $c_{\delta_t}^{(i)}$ in Eq.~\eqref{eq:autocorrelation}. Following the example of Eq.~\eqref{eq:transfer_operator_MPO}, this incorporates the repeated, joined unitary evolution [gray gates] and appropriate projections on the ancillary degrees of freedom [white and red diamond for `0' and `1', respectively]. Note that we use double lines to distinguish the MPO evolution in the space of density matrices from the MPS evolution of a pure state in Fig.~\ref{fig:fig_s1}(a).}
    \label{fig:fig_s2}
\end{figure}

Taking the probability in Eq.~\eqref{eq:probability_inactive_regions} and the autocorrelation function in Eq.~\eqref{eq:autocorrelation} as an example, we show their corresponding tensor network representations in panels (a) and (b) of Fig.~\ref{fig:fig_s2}, respectively. In both panels, the dynamics is implemented in a folded space, where two layers similar to the one in Fig.~\ref{fig:fig_s1}(a) are combined to represent the evolution in the space of density matrices. 
Representing the system-ancilla state as a vectorized Matrix Product Operator~(MPO), the unitary evolution, shown as connected gray gates in Fig.~\ref{fig:fig_s2}, can still be obtained via the Trotter-Suzuki decomposition in analogy with the procedure outlined in the previous section. Here, we set a maximum bond dimension $\chi$ in the SVDs of the TEBD algorithm.  
This step is followed by applying single-site projectors on state $\ket{0_\mathrm{A}}$ or $\ket{1_\mathrm{A}}$ [see $\ProjZeroMPO$ or $\ProjOneMPO$ in Fig.~\ref{fig:fig_s2}], when calculating the probability of an inactive cluster or the autocorrelation function in Eq.~\eqref{eq:autocorrelation}, respectively.
Finally, we trace over all ancillas before resetting them to the state $\ket{\mathbf{0}_\mathrm{A}}$ [see $\TraceAndReset$ in Fig.~\ref{fig:fig_s2}] to prepare for the next time step. \\ 

Let us also remark that the tensor network diagram representation does not stipulate that contractions must be performed in the forward direction, as described above. 
Taking inspiration from Refs.~\cite{Barthel2013,Gillman2019,FriasPerez2022a}, we can use the backward or two-way evolution to reduce the computational cost as measured by the required bond dimension and the number of individual contractions to obtain the same level of accuracy (see Tab.~\ref{tab:TN_description} for the individual specifications). Furthermore, to avoid instabilities due to small numerical values, we always compute and store the multiplicative increment, e.g., $p_{\ell \times (\tau + 1)} / p_{\ell \times \tau}$, and only later access the full quantity as cumulative product.\\

\begin{figure}[ht]
    \centering
    \includegraphics{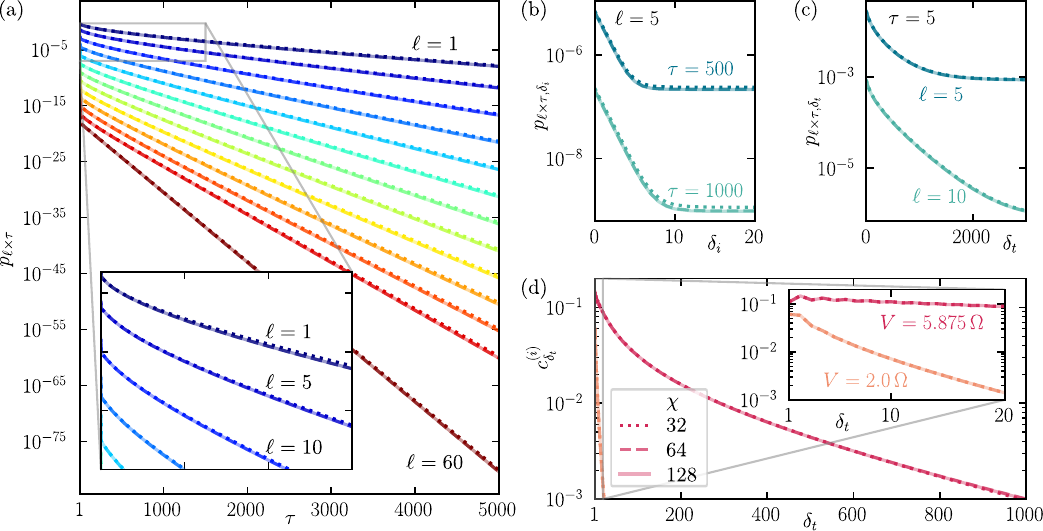}
    \caption{\textbf{Convergence checks of average state calculations.} We compare results obtained for different bond dimensions $\chi = 32, 64, 128$ as dotted, dashed and solid lines with fading opacity. (a)~Probability $p_{\ell \times \tau}$ of inactive space-time regions with fixed spatial size $\ell \in \{1, 5, 10, ..., 60\}$ as a function of the temporal extension $\tau$ entering Figs.~\ref{fig:fig2}~and~\ref{fig:fig3}. (b,c)~Probability $p_{\ell \times \tau, \delta_i}$ and $p_{\ell \times \tau, \delta_t}$ of inactive regions separated by $\delta_i$  $\delta_t$ considered in Fig.~\ref{fig:fig3}. (c)~Autocorrelation function for the ideal case as a function of the time difference $\delta_t$ shown in Fig.~\ref{fig:fig4}. If not stated otherwise, we consider $L=60$, $\Omega \Delta t = 1.25$, $V=5.875\,\Omega$, $\gamma = 3\,\Omega$ and $M=10$.}
    \label{fig:fig_s3}
\end{figure}

Finally, we verify that the chosen bond dimension of $\chi  = 64$ is sufficient to capture the quantities presented in Figs.~\ref{fig:fig2}-\ref{fig:fig4}. We therefore repeat all simulations with both decreased and increased accuracy, and display the resulting probabilities or the autocorrelation function over time in Fig.~\ref{fig:fig_s3}. Here, we observe that simulations with a bond dimension of $\chi = 32$ show minor differences, while the agreement between $\chi = 64$ and $\chi = 128$ signals the convergence of the results.


\section{Further discussion on inactive clusters}
\begin{figure}[t]
    \centering
    \includegraphics{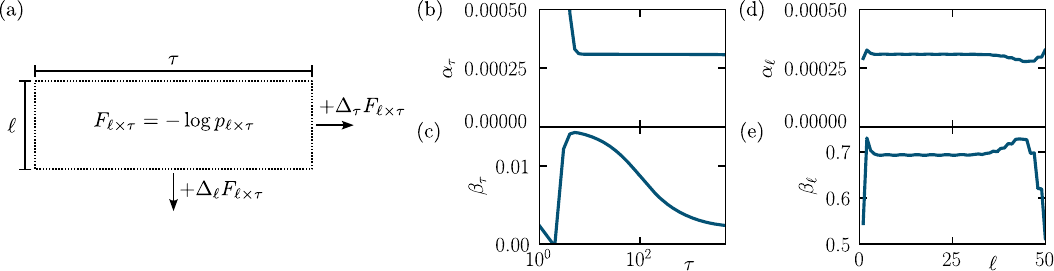}
    \caption{\textbf{Inactive space-time cluster.} (a)~Graphical representation of the spatial and temporal interface tensions of inactive cluster. 
    (b,c)~Numerical coefficients $\alpha_\tau$ and $\beta_\tau$ as extracted from linear fitting of $\Delta_\tau F_{\ell \times \tau}$ in Eq.~\eqref{eq:scaling_function} within the range of $10 \leq \ell \leq 40$. (d,e)~Numerical coefficients $\alpha_\ell$ and $\beta_\ell$ as extracted from linear fitting of $\Delta_\ell F_{\ell \times \tau}$ according to Eq.~\eqref{eq:scaling_function_space} in the interval specified by $2000 \leq \tau \leq 3000$. Here, $L=60$, $\Omega \Delta t = 1.25$, $\gamma = 3\,\Omega$, $V=5.875\,\Omega$ with $M = 10$ and $\chi = 64$.}
    \label{fig:fig_s4}
\end{figure}

In this section, we provide further discussions of the inactive clusters and their corresponding free energies. In particular, we compare the associated interface tensions $\Delta_{\ell} F_{\ell \times \tau}$ and $\Delta_{\tau} F_{\ell \times \tau}$ depicted in Fig.~\ref{fig:fig_s4}(a), which represent the free-energy cost of extending the cluster in either space or time.\\

We start by recalling the definition of the dynamical free energy in Eq.~\eqref{eq:dynamical_free_energy},  which is expressed  in terms of the probability of finding an inactive region of size $\ell \times \tau$. This probability can be computed using the procedure outlined in the previous section. Taking inspiration from  Refs.~\cite{Katira2018,Klobas2024,DeFazio2024}, we study the scaling behavior of the dynamical free energies by making the following ansatz
\begin{align}
    F_{\ell \times \tau} = \alpha_{\ell, \tau} \ell \tau + \beta_\tau \tau + \beta_\ell \ell + \mathrm{const.} \, .
    \label{eq:ansatz_dynamical_free_energy}
\end{align}
The coefficients $\alpha$ and $\beta$ are evaluated numerically and control the contributions of the area $\ell \tau$, and the perimeter $2(\ell + \tau)$ of the considered inactive space-time cluster, respectively. 
Note however that, as indicated by their corresponding subscripts, these coefficients may depend on $\ell$ and $\tau$, and explicitly allow for anisotropy in space and time. \\

Eq.~\eqref{eq:ansatz_dynamical_free_energy} is consistent with the ansatz for the temporal interface tension in Eq.~\eqref{eq:scaling_function}, whose numerical parameters are shown in the central panels of Fig.~\ref{fig:fig_s4}. We notice that the crossover from (I) to (II) observed in Fig.~\ref{fig:fig2}(c) occurs due to the decrease of $\alpha_\tau$ and a simultaneous increase of $\beta_\tau$. On the other hand, we find that $\beta_\tau$ then decreases again, making it, in principle, easier to maintain a inactive space-time region while reestablishing a more dominant role for the area scaling with $\alpha_\tau$, as seen in the crossover from (II) to (III) in Fig.~\ref{fig:fig2}(c). \\

Similarly, an ansatz can be formulated for the spatial interface tension, defined as the increment in the dynamical free energy obtained by extending the inactive space-time cluster by a unit of space, which reads
\begin{align}
    \Delta_\ell F_{\ell \times \tau} = F_{(\ell + 1) \times \tau} - F_{\ell \times \tau} = \alpha_\ell \tau + \beta_\ell \, .
    \label{eq:scaling_function_space}
\end{align}
Numerical values of the parameters $\alpha_\ell$ and $\beta_\ell$ \, are shown in Fig.~\ref{fig:fig_s4}, right panels. 
We observe that the inferred values of $\alpha_\ell$ in panel~(d) agree well with those of $\alpha_\tau$ in panel~(b) over a large range of cluster sizes, further confirming the applicability of the ansatz in Eq.~\eqref{eq:ansatz_dynamical_free_energy}.
Furthermore, we note that the perimeter-related contribution given by $\beta_\ell$ in panel~(e) is significantly larger than the counterpart in time shown in panel~(c). This indicates that it is more favorable to extend clusters in time rather than in space. On the one hand, this difference is already reflected in the clusters with long temporal extents observed in Fig.~\ref{fig:fig1}(d). 
On the other hand, it explains why correlations persist longer in time than in space [cf.~Fig.~\ref{fig:fig3}].
Finally, we note that the deviation from approximately constant values of $\alpha_\ell$ and $\beta_\ell$ for $\ell \gtrsim 40$ indicates the presence of boundary effects. 


\section{Tuneable dynamical phase coexistence}
\label{sec:tunable_dynamical_phase_coexistence}
In this section, we investigate the dependence of the dynamical phase coexistence, as seen in Fig.~\ref{fig:fig1}(c,d), on the atom-atom interaction strength $V$ [cf.~Eq.~\eqref{eq:collision_model_hamiltonian}].\\

\begin{figure}[ht]
    \centering
    \includegraphics{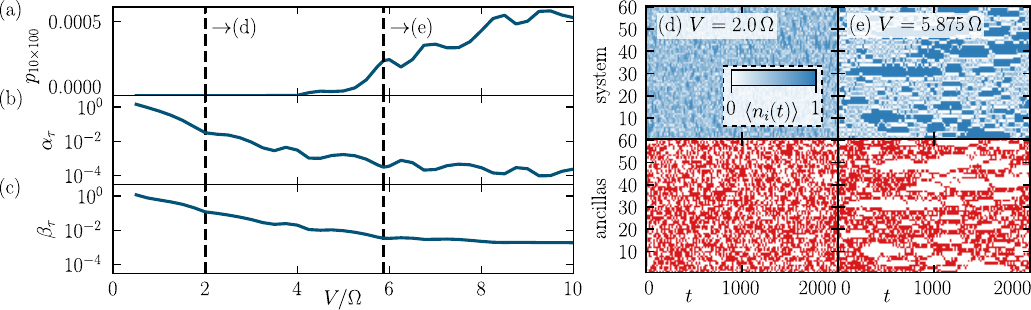}
    \caption{\textbf{Tuneable dynamical phase coexistence}. (a)~Probability of $10\!\times\!100$-sized inactive space-time regions for atom-atom interaction strengths $V$. (b,c)~Area- and perimeter-scaling coefficients $\alpha_\tau$ and $\beta_\tau$. We show the coefficients extracted from fitting the temporal interface tension, $\Delta_\tau F_{\ell \times \tau}$, at $\ell \in \{10, 20, 30, 40\}$ and $\tau = 1000$ according to Eq.~\eqref{eq:scaling_function}. (d,e)~Stochastic realizations and corresponding space-time resolved quantum trajectory for $V=2\,\Omega$ and $V=5.875\,\Omega$ [see vertical dashed lines in panels~(a-c)]. We consider $L=60$, $\gamma = 3\,\Omega$, $\Omega \Delta t = 1.25$ , $M=10$ with $\chi = 64$ (a-c) or $\epsilon = 10^{-10}$ (d,e) respectively.}
    \label{fig:fig_s5}
\end{figure}

Figure~\ref{fig:fig_s5}(a) shows how the probability of finding $\ell \times \tau$-sized inactive cluster, i.e., for $\ell= 10$ and $\tau=100$, varies with the interaction strength $V$. We observe that this probability is distinguishable from zero for $V \gtrsim 4\,\Omega$, where $\Omega$ is the Rabi frequency. 
We observe that this probability generally increases with $V$, while displaying oscillatory behavior. To understand this behavior, we need to consider two contributions. On the one hand, the dynamical phase coexistence becomes more pronounced as $V$ increases, factually freezing the dynamics of adjacent atoms in the excited state $\ket{1_\mathrm{S}}$ for $V \to \infty$ \cite{Lesanovsky2012a,Turner2018}. On the other hand, the superimposed oscillatory behavior can be understood by recalling that the discrete-time quantum dynamics can display resonances, which would not be present in the continuous-time generator, i.e., $H_\mathrm{CM}$ in Eq.~\eqref{eq:collision_model_hamiltonian} \cite{Cech2025,Wald2025}. We note that the increase in the probability in panel~(a) is directly connected to the decrease of several orders of magnitude in the coefficients $\alpha_\tau$ and $\beta_\tau$ in panels~(b,c).\\

Finally, we show two stochastic realizations alongside their respective space-time resolved quantum trajectories in Figs.~\ref{fig:fig_s5}(d,e). For the chosen values of $V=2\,\Omega$ and $V=5.875\,\Omega$, only the trajectory for $V= 5.875\,\Omega$ displays a large-scale coexistence of inactive and active space-time regions, thus confirming the expectation from panels~(a-c).


\section{Details on empirical quantities of quantum trajectories}
In this section, we discuss the details of evaluating the simulated trajectories empirically in order to extract the probabilities and the autocorrelation function shown as symbols in Fig.~\ref{fig:fig2} and Fig.~\ref{fig:fig4}, respectively. In particular, we emphasize and demonstrate that our analysis does not require any postselection, which would be necessary when characterizing the properties of a quantum state in a stochastic realization. Finally, we investigate the influence of averaging over both space and time in the presence of experimental imperfections. 
\\

Following the approach outlined in Ref.~\cite{Cech2025}, here applied to larger system sizes, we employ the concept of time-integrated observables on individual trajectories to estimate the probabilities of inactive regions and the autocorrelation function in two stages. 
To illustrate this concept, consider the calculations performed to obtain the empirical estimate for the autocorrelation function of the ancilla outcomes in Eq.~\eqref{eq:autocorrelation}. For a single quantum trajectory $\eta(T) = [\mathbf{k}(t)]_{t=1}^T$ with $\mathbf{k} = (k_1, ..., k_L)$, we first calculate 
\begin{align}
    \label{eq:autocorrelation_function_empirically}
    c_{\delta_t}[\eta(T)] = \frac{1}{L' T'} \mathcal{O}_{\delta_t}[\eta(T)] - \Big( \frac{1}{L' T} \mathcal{O}_{0}[\eta(T)] \Big)^2 \, .
\end{align}
Here, we introduced time-integrated observables as a generalization of the average number of `1' measurement outcomes in Eq.~\eqref{eq:average_activity} as
\begin{align}
    \label{eq:time_integrated_observable}
    \mathcal{O}_{\delta_t}[\eta(T)] = \sum_{{t = 1+\delta_t}}^{\mathclap{T}}  \sum_{{i=1 + i_0}}^{\mathclap{L - i_0}} k_{i}(t) k_i(t - \delta_t) \, ,
\end{align}
and the respective extensivities, $L' = L - 2 i_0$ and $T' = T - \delta_t$, of the included sums. 
By restricting ourselves to ancilla outcomes that are, if not stated otherwise, $i_0=5$ sites away from the boundaries, we avoid boundary effects while simultaneously increasing the statistical accuracy as compared to evaluating a single ancilla site $i$. 
We then expect that the long-term average in Eq.~\eqref{eq:autocorrelation_function_empirically} accesses the stationary behavior of the autocorrelation function of quantum trajectories in Eq.~\eqref{eq:autocorrelation}.
After averaging over space and time, we calculate the mean of several simulated trajectories to determine their empirical value. We then estimate the statistical uncertainty to be twice the standard error. 
Considering the concrete example of inactive regions in Fig.~\ref{fig:fig2}(a), $20$ quantum trajectories with $100\,000$ time steps each are sufficient to capture the probabilities of inactive regions well down to $10^{-4}$.

\begin{figure}[ht]
    \centering
    \includegraphics{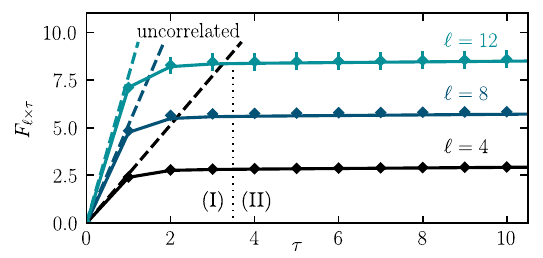}
    \caption{\textbf{Statistics of inactive space-time clusters with experimentally accessible trajectory ensembles.} We show the dynamical free energy of finding $\ell \times \tau$-sized inactive clusters as a function of the temporal extension $\tau$ for fixed spatial extensions $\ell = 4, 8, 12$. Other than in Fig.~\ref{fig:fig2}(a), we here evaluate the empirical probabilites (symbols with errorbars indicating the statistical uncertainty) obtained from $400$ simulated trajectories of 20 time steps each. Dashed lines represent the predictions for uncorrelated measurement outcomes and serve as a guide for the eye. We consider $L=60$, $\Omega \Delta t = 1.25$, $V=5.875\,\Omega$, $\gamma = 3\,\Omega$ and $M=10$ with $\chi = 64$ (average state calculations) and $\epsilon = 10^{-8}$ (quantum trajectories), respectively. Note that the simulated trajectories are not subject to disordered couplings or bit-flip errors, both of which are considered in Fig.~\ref{fig:fig_s7}.}
    \label{fig:fig_s6}
\end{figure}

In a potential experimental implementation, similar evaluations could also benefit directly from the knowledge of the stationary state, which can be accessed by initializing the system in random product states in the computational basis. This allows long trajectories to be replaced by shorter ones with different initial states, while still sampling from the stationary distribution. To illustrate this alternative approach for Fig.~\ref{fig:fig2}(a), we consider the empirical evaluation of 400 quantum trajectories, each 20 time steps long, in Fig.~\ref{fig:fig_s6}. We observe that this is again sufficient to distinguish between area- and perimeter-dominated scaling regimes, and to capture the crossover between them.

\begin{figure}[ht]
    \centering
    \includegraphics{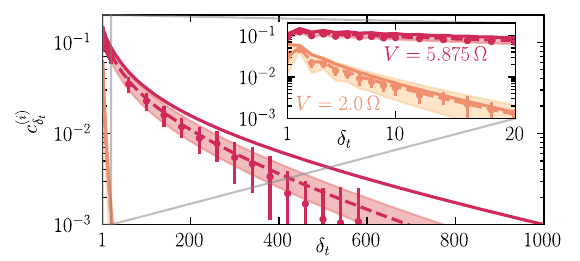}
    \caption{\textbf{Alternative view on the autocorrelation function of space-time records.} Similar to Fig.~\ref{fig:fig4}, we compare the ideal dynamics (solid lines) to observations made under experimental imperfections. For the latter, we consider ten disorder sets of disordered couplings, i.e., $V$ and $\gamma$ locally varied within a range of up to $\pm \Omega$. For all of them, we simulate both a quantum trajectory over $100\,000$ time steps (symbols, incorporating also $p_\mathrm{err} = 2\,\%$ of bit-flip errors) and appropriately conditioned average state calculations (dashed lines with shaded regions indicating the statistical uncertainties). Note that in contrast to Fig.~\ref{fig:fig4}, we use $i_0 = L/2 - 1 = 29$ in Eqs.~(\ref{eq:autocorrelation_function_empirically},\ref{eq:time_integrated_observable}) and therefore evaluate only the most central ancilla outcomes. Throughout, we set $L=60$, $\gamma = 3\,\Omega$, $\Omega \Delta t = 1.25$, $M=10$ with $\chi = 64$ (average state calculations) or $\epsilon = 10^{-8}$ (quantum trajectories) respectively.}
    \label{fig:fig_s7}
\end{figure}

To highlight also the effect of the spatial average in the empirical evaluation, we revisit the autocorrelation function in the presence of experimental imperfections. In contrast to Fig.~\ref{fig:fig4}, where the square symbols are extracted from a large number of ancilla measurement outcomes in the bulk, we now focus only on the central ones. In Fig.~\ref{fig:fig_s7}, we then observe two important aspects of this modification. Firstly, as expected, we notice an increase in statistical uncertainty. Furthermore, the autocorrelation function with experimental imperfections, which considers only the most central ancillas, decreases slightly faster for $V = 5.875\,\Omega$ than was previously observed in Fig.~\ref{fig:fig4}. 
In fact, the local maximum of the probability for a mesoscopic inactive cluster, as observed around $V = 5.875\, \Omega$ in Fig.~\ref{fig:fig_s5}(a), suggests that changes to the parameters of the Hamiltonian in Eq.~\eqref{eq:collision_model_hamiltonian} would have a diminishing impact on the signatures of the dynamical phase coexistence. However, when we compare Fig.~\ref{fig:fig4} with Fig.~\ref{fig:fig_s7}, we can see that incorporating ancilla measurement outcomes over a larger spatial region in the quantum trajectory effectively increases the likelihood of finding long-lived correlations that resemble the ones in the ideal case.

\end{document}